\input{aipcheck}

\documentclass[
    ,final            % use final for the camera ready runs
  ]
  {aipproc}

\layoutstyle{8x11double}

\usepackage{amsmath,amssymb}

\begin{document}
\newcommand{\la}{\langle}
\newcommand{\ra}{\rangle}
\newcommand{\nn}{\nonumber}
\newcommand{\ep}{\varepsilon}
\newcommand{\vP}{{\mathbf{P}}}
\newcommand{\vp}{\mathbf{p}}
\title{Scalar-Pseudoscalar scattering and pseudoscalar resonances}


\author{M. Albaladejo}{
  address={Departamento de F\'isica, Univerisidad de Murcia, E-30071, Murcia, Spain}
}

\author{J.A. Oller}{
}

\author{L. Roca}{
}

\begin{abstract}
The interactions between the $f_0(980)$ and $a_0(980)$  scalar resonances and the  
lightest pseudoscalar mesons are studied. We first obtain the interacting kernels, without including any \textit{ad hoc} free parameter, because the lightest scalar resonances are dynamically generated. These kernels are unitarized, giving the final amplitudes, which generate pseudoscalar resonances, associated with the $K(1460)$, $\pi(1300)$, $\pi(1800)$, $\eta(1475)$ and $X(1835)$. We also consider the exotic channels with $I=3/2$ and $I^G=1^+$ quantum numbers. The former could be also resonant in agreement with a previous prediction.  
\end{abstract}

\maketitle

%%%%%%%%%%%%%%%%%%%%%%%%%%%%%%%%%%%%%%%%%%%%
%% MAINMATTER
%%%%%%%%%%%%%%%%%%%%%%%%%%%%%%%%%%%%%%%%%%%%

\section{Introduction}

Chiral symmetry imposes strong constraints to the interactions of the lightest pseudoscalar mesons ($\pi$, $K$, $\eta$, $\eta'$). For the isospin ($I$) 0,~1 and 1/2 the scattering of the pseudoscalars in S-wave is strong enough to  generate dynamically the lightest scalar resonances, namely, the $f_0(980)$, $a_0(980)$, $\kappa$ and $\sigma$, as shown in Ref.~\cite{chptja}. One can make use of the tightly constrained interactions among the lightest pseudoscalars in order to work out approximately the scattering between them and scalar resonances. We focus here on the relatively narrow $f_0(980)$ and $a_0(980)$ resonances and consider their interactions with the pseudoscalars $\pi$, $K$, $\eta$ and $\eta'$. These interactions are shown to be strong enough to generate dynamically new pseudoscalar resonances with $J^{PC}=0^{-+}$, with a mass larger than 1~GeV, typically following the relevant thresholds.

We report here on our recent paper \cite{citpseudo} dedicated to the interesting problem of the excited pseudoscalar mesons above 1~GeV. These resonances are not typically well-known \cite{pdg}. In $I=1/2$ one has the $K(1460)$ and $K(1630)$ resonances. The $I=1$ resonances $\pi(1300)$, $\pi(1800)$ are somewhat better known. Special mention deserves the $I=0$ channel where the $\eta(1295)$, $\eta(1405)$ and $\eta(1475)$ have been object of an intense study \cite{masoni_rev}. Experimentally it has been established that while the $\eta(1405)$ decays mainly to $a_0\pi$ the $\eta(1475)$ does so to $K^*\bar{K}+c.c$. Refs.~\cite{pdg,masoni_rev} favor the interpretation of considering the $\eta(1295)$ and $\eta(1475)$ as ideally mixed states of the same nonet of pseudoscalar resonances, the other members being the $\pi(1300)$ and  $K(1460)$, forming the first radial excitation of the lightest pseudoscalars. The $\eta(1405)$ would then be an extra state, that, if interpreted as a glueball in QCD, it would be in conflict with present results from lattice QCD, predicting the lowest mass for the pseudoscalar glueball at around 2.4~GeV. The previous whole picture has been challenged in Ref.~\cite{klempt} that questions the existence of the $\eta(1295)$ and argue that only one $\eta(1440)$ exists in the 1.4-1.5 GeV region.

Recently, the BES Collaboration \cite{besx1835} has observed the new resonance $X(1835)$ with quantum numbers favored as a pseudoscalar $0^{-+}$ resonance.

\section{Formalism}
\label{sec2}

\begin{figure}\centering
\includegraphics[height=3cm,keepaspectratio]{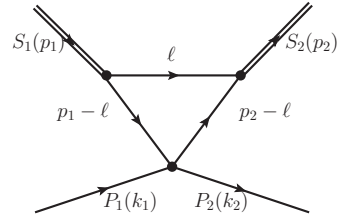}
\caption{Triangle loop for calculating the interacting kernel of $S_1(p_1) P_1(k_1) \to S_2(p_2) P_2(k_2)$, where the four-momentum for each particle is given between brackets. $S_{1,2}$ represents the initial, final scalar resonances and similarly for $P_{1,2}$ regarding
the pseudoscalar mesons. 
\label{fig:tri}}
\end{figure}

Our approach is based on the triangle diagram of Fig.~\ref{fig:tri}, where an initial scalar ($S_1$) and pseudoscalar ($P_1$) go to a final pair ($S_2$, $P_2$), with the internal lines being kaons. This diagram is enhanced because the masses of both the $f_0(980)$ and  $a_0(980)$ resonances are very close to the $K\bar{K}$ threshold, as discussed in detail in Ref.~\cite{alva}. The couplings of the scalar resonances to two kaons are determined by studying the meson-meson interaction \cite{citpseudo}, similarly as in Ref.~\cite{chptja}.
 
Let us indicate  by $P$ the total four-momentum $P=p_1+k_1=p_2+k_2$ in Fig.~\ref{fig:tri}.
  This diagram is given by $g_1 g_2 L_K$, being $g_i$ the coupling of the initial and final scalar resonance to a $K\bar{K}$ pair, and $L_K$ corresponds to:
 \begin{align}
 L_K &= i \int\frac{d^4\ell}{(2\pi)^4}\frac{T((P-\ell)^2)}{(\ell^2-m_K^2+i\ep)((p_1-\ell)^2-m_K^2+i\ep)} \nn \\ & \times \frac{1}{((p_2-\ell)^2-m_K^2+i\ep)}~.
 \label{tl.ref}
 \end{align}
  Here, $T((P-\ell)^2)$ represents the interaction amplitude between the kaons with the external pseudoscalars, see Ref.~\cite{citpseudo} for details. These amplitudes contain the poles corresponding to the scalar resonances $\sigma$, $\kappa$, $f_0(980)$, $a_0(980)$ and other poles in the region around 1.4~GeV \cite{chptja,miguel}. To work out the dependence of $T$ on the integration variable, we write the dispersion relation satisfied by $T(q^2)$,
\begin{align}
T(q^2)&=T(s_A)+\sum_i\frac{q^2-s_A}{q^2-s_i}\frac{\hbox{Res}_i}{s_i-s_A}
+ \nn \\ + & \frac{q^2-s_A}{\pi}\int_{s_{th}}^\infty ds'\frac{\hbox{Im}T(s')}{(s'-q^2)(s'-s_A)}~.
\label{dis.rel}
\end{align} 
One subtraction at $s_A$ is taken. There are also present 
 poles deep in the $q^2$-complex plane located at $s_i$ whose residues are Res$_i$. Inserting Eq.~\eqref{dis.rel} into Eq.~\eqref{tl.ref}, it results
\begin{align}
L_K&=\left(T(s_A)+\sum\frac{\hbox{Res}_i}{s_i-s_A}\right)C_3
+\sum_i C_4(s_i) \hbox{Res}_i \nn\\
&-\frac{1}{\pi}\int_{s_{th}}^\infty ds' \hbox{Im}T(s') 
\left[\frac{C_3}{s'-s_A}+C_4(s') \right]~\text{,}
\label{tl.dis}
\end{align} 
where we have introduced the three- and four-point Green functions $C_3$ and $C_4(M_4^2)$, that must be projected into S-wave. Once this is done, eq.~\eqref{tl.dis} can still be used but with $C_3$ and $C_4(M_4^2)$ projected \cite{citpseudo}.
 
\begin{table}[h!]
\begin{tabular}{ccc}
\hline
  \tablehead{1}{c}{b}{$I^G$} 
& \tablehead{1}{c}{b}{Channels}
& \tablehead{1}{c}{b}{Resonances} \\
\hline
$0^+$         & $f_0 \eta$, $f_0 \eta'$, $a_0\pi$ & $\eta(1295)$, $\eta(1405)$,\\ & & $\eta(1475)$, $X(1835)$\\
$\frac{1}{2}$ & $f_0 K$, $a_0 K$ & $K(1460)$, $K(1630)$\\
$1^-$         & $f_0\pi$, $a_0 \eta$, $a_0\eta'$ & $\pi(1300)$, $\pi(1800)$\\
$\frac{3}{2}$ & $a_0 K$ & $K\bar{K}K$ (Longacre \cite{longacre})\\
$1^+$         & $a_0 \pi$ & -- \\
\hline
\end{tabular}
\caption{Different $I^G$ sectors, coupled channels and resonances \cite{pdg}.\label{tab:chan}}
\end{table}

Eq.~\eqref{tl.dis} is our basic equation for evaluating the interaction kernels, giving rise to the interaction kernels $T_L(i\to j)$, where $i$, $j$ refer to the different scalar-pseudoscalar channels involved, as seen in Table \ref{tab:chan}. For each $I^G$ sector, we join in a symmetric matrix ${\cal T}_{IG}$ the different $T_L(i\to j)$. In order to resum the unitarity loops and obtain  the final S-wave T-matrix, $T_{IG}$, we use the equation (see e.g. \cite{citpseudo,miguel} and references therein)
 \begin{align}
T_{IG}&=\left[I+{\cal T}_{IG}\cdot g_{IG}(s)\right]^{-1}\cdot {\cal T}_{IG}~.
\label{t.uni}
\end{align}
Here, $g_{IG}(s)$ is a diagonal matrix whose elements are the scalar unitarity loop function with a scalar-pseudoscalar intermediate state. It has a subtraction constant ($a_1$) restricted to have natural values, so that the associated unitarity scale $4\pi f_\pi/\sqrt{|a_1|}$ becomes not too small. The sign of $a_1$ is required to be negative so that resonances could be generated when the interaction kernel is positive (attractive). Then, from these two conditions, the resulting resonances might be qualified as dynamically generated due to the iteration of the unitarity loops. 

\section{Results}\label{sec3}
\begin{figure}[h!]
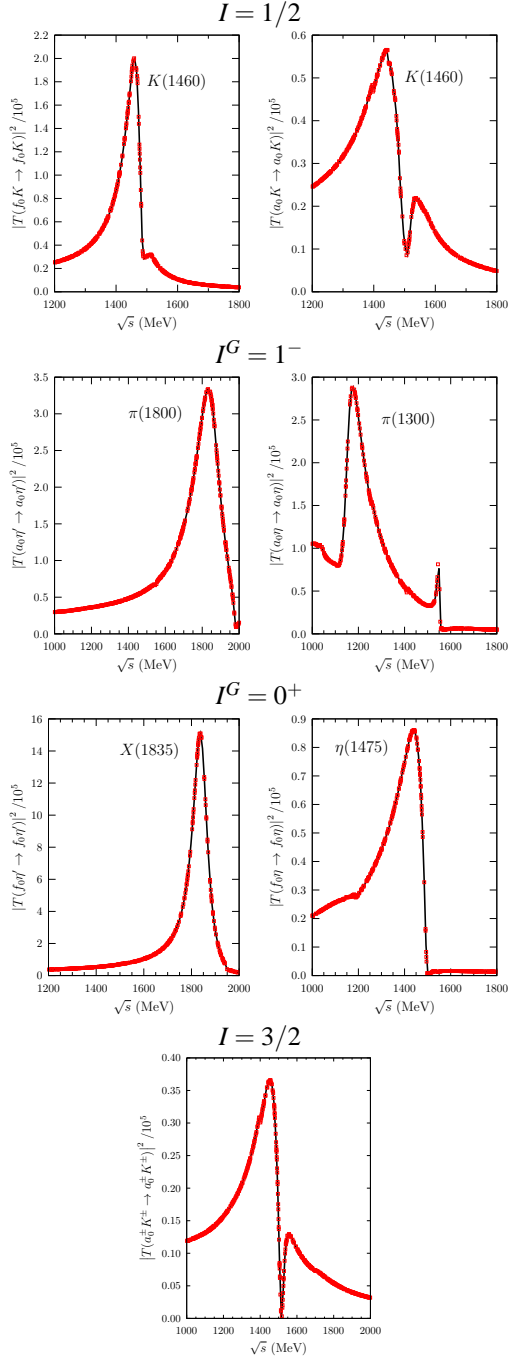

\begin{tabular}{c}
$I=1/2$ \\ \includegraphics[height=4cm,keepaspectratio]{I12_QCHSIX.eps} \\
$I^G=1^-$   \\ \includegraphics[height=4cm,keepaspectratio]{I1_QCHSIX.eps} \\
$I^G=0^+$   \\ \includegraphics[height=4cm,keepaspectratio]{I0_QCHSIX.eps} \\
$I=3/2$ \\ \includegraphics[height=4cm,keepaspectratio]{I32_QCHSIX.eps}
\end{tabular}
\caption{Modulus squared of the different scalar-pseudoscalar amplitudes discussed in the text.
\label{fig:amplis}}
\end{figure}

In this section we show the results that follow by applying Eq.~\eqref{t.uni} to the different 
$I^G$ sectors. The modulus squared of the amplitudes are shown in Fig.~\ref{fig:amplis}. The resonances found in our work are given in Table~\ref{tab:res}.
\begin{table}[h!]
\begin{tabular}{lccl}
\hline
  \tablehead{1}{c}{b}{Resonance} 
& \tablehead{1}{c}{b}{$I^{G}$}
& \tablehead{1}{c}{b}{Width (MeV)} 
& \tablehead{1}{c}{b}{Comments} \\
\hline
$K(1460)$ & $\frac{1}{2}$ & $\Gamma\gtrsim 100$ & $|g_{f_0 K}/g_{a_0 K}|\simeq 1.4$ \\
$\pi(1800)$ & $1^{-}$ & $\Gamma\simeq 200$  & $a_0 \eta'$ elastic \\
$\pi(1300)$ & $1^{-}$ & $\Gamma\gtrsim 200$ & $a_0\pi$, $f_0\eta$ coupled \\
$X(1835)$ & $0^+$ & $\Gamma\simeq 70$       & $f_0\eta'$ elastic \\
$\eta(1475)$ & $0^+$ & $\Gamma\simeq 150$   & $f_0\eta$ elastic\\
Exotic & $\frac{3}{2}$ & $\Gamma\simeq 200$   & $a_0 K$ threshold \\
\hline
\end{tabular}
\caption{Summary of the resonances obtained in our study.
\label{tab:res}}
\end{table}
\paragraph{${I=1/2}$ sector} We obtain in $f_0 K \to f_0 K$ a clear resonant peak with its maximum at $1460$~MeV for $a_1$ around $-0.5$, that corresponds to the nominal mass of the $K(1460)$ resonance. The  visual width of the peak is around 100~MeV, although it appears wider in $a_0 K\to a_0 K$ scattering. The PDG \cite{pdg} refers a larger width of around 250~MeV. One has to take into account that the channel $K^*(892)\pi$ is not included and it seems to couple strongly with the $K(1460)$ resonance. From the relative sizes of the peaks one infers that the $K(1460)$ couples more strongly to $f_0 K$ than to  $a_0 K$.
 
\paragraph{$I=1$ sector} We find that the $a_0\eta'$ channel is almost elastic due to its much higher threshold \cite{citpseudo}. It gives rise to a strong resonant signal around 1.8~GeV that could be associated with the  $\pi(1800)$ resonance. Taking $a_1$ for $a_0 \eta'$ around $-1.3$, we get the mass given in the PDG ($1816\pm 14$~MeV). The visual width of the peak is around 200~MeV (PDG quotes $208\pm 12$~MeV). The other two channels couple quite strongly and give rise to an enhancement between 1.2-1.4~GeV, that we identify with the $\pi(1300)$.

\paragraph{$I=0$ sector} As in $I=1$, the $f_0\eta'$ channel decouples from the other two channels. For $a_1 \simeq -1.2$ one obtains a resonance of the $a_0\eta'$ channel with a mass of 1835~MeV and a width around 70~MeV, in agreement with PDG for the $X(1835)$, $1833.7\pm 6.1\pm 2.7$~MeV and $67.7\pm 20.3\pm 7.7$~MeV, respectively. For the other two channels, $a_0\pi$ and  $f_0\eta$, we obtain a clear resonance with a mass around 1.45~GeV not coupled to $a_0\pi$. As $\eta(1405)$ couples mostly to $a_0\pi$ we identify our resonance with the $\eta(1475)$. The width is around 150~MeV (PDG gives $85\pm 9$~MeV). It is also known that the $\eta(1475)$ couples strongly to $K^*(892)\bar{K}+c.c$, not included in our study, which would modify its shape. No resonance  around the mass of the $\eta(1295)$ is observed.

\paragraph{Exotic channels} In $I=3/2$ we find an exotic resonance at around 1.4 GeV for $|a_1| \lesssim 1.5$. This confirms the predictions of Longacre \cite{longacre} that studied the $K\bar{K}\pi$ and $K\bar{K}K$ systems, concluding that the $I=3/2$ $J^{P}=0^{-}$  $K \bar{K}K$ system was resonant around its threshold.

In the other exotic channel ($I^G = 1^+$), a resonant behavior depends on the value of $a_1$. For $|a_1|\lesssim 1$ the resonant signal is weak, and that value is finally taken because no such resonance has  been found experimentally.   

\vspace{0.5cm}
In summary, we have developed a novel approach for studying the scalar-pseudoscalar interactions that also sheds new light on the pseudoscalar resonances with mass above 1 GeV.
\begin{theacknowledgments}
This work has been partially funded by the MEC grant FPA2007-6277 and Fundaci\'on S\'eneca grant 11871/PI/09. M.A. acknowledges the Fundaci\'on S\'eneca for the FPI grant 13310/FPI/09.
\end{theacknowledgments}



\bibliographystyle{aipproc}   % if natbib is available

\end{document}